\documentclass{article}
\usepackage{graphicx}
\usepackage[numbers]{natbib}
\usepackage{doi}

\begin{document}

\title{Modeling network technology deployment rates with different network models}
\author{Yoo Chung}
\maketitle

\begin{abstract}
  To understand the factors that encourage the deployment of a new networking technology, we must be able to model how such technology gets deployed.  We investigate how network structure influences deployment with a simple deployment model and different network models through computer simulations.  The results indicate that a realistic model of networking technology deployment should take network structure into account.
\end{abstract}

\section{Introduction}
\label{sec:introduction}

On January 31, 2011, the Internet Assigned Numbers Authority allocated the last free blocks of IPv4 addresses to the five Regional Internet Registries~\cite{w:ipv4-free-pool-depleted}.  It will not be long before even the Regional Internet Registries run out of their own IPv4 addresses to give out.  This is the sort of news which gives urgency to the need for IPv6 to replace IPv4 for its greatly expanded address space.  And yet IPv6 deployment rates, while they are steadily climbing, remain extremely low~\cite{karpilovsky:pam2009,colitti:pam2010}.

Looking to the distant future, clean-slate design approaches are being proposed to solve problems that seem to be insurmountable or too messy to solve with the design of the Internet today~\cite{feldmann:sigcomm2007,roberts:aot2009}.  Perhaps the technical community may go further than just using a clean-slate design approach for generating new ideas and converge on a clean-slate design for a new Internet that can solve all of the problems with the present Internet.  But if IPv6 can hardly get deployed even when there is a clear perceived need for it today, how can we expect this clean-slate Internet to replace the by-then obsolete Internet?

In order to figure out how to design new networking technologies so that they are quickly adopted, we need to understand the factors that are behind the spread of such technologies.  To understand these factors, we need to be able to model the deployment of these technologies to a reasonable degree of accuracy.

This work does not attempt to define a comprehensive model for the deployment of new networking technologies, nor does it try to suggest what policies or design principles may promote the deployment of such technologies.  Instead, this work focuses on how different \emph{network structure} may influence the spread of a new networking technology throughout a network, an understanding of which may be important in the design of a realistic model.  Using computer simulations, we investigate how different network models influence deployment rates.  Other work has focused on the deployment of specific networking technologies~\cite{chan:sigcomm2006} or on the role of ``converters'' which provide compatibility with an old technology in the deployment of a new technology~\cite{joseph:conext2007,sen:ton2010}.

The rest of the paper is structured as follows.  Section~\ref{sec:infection} describes the deployment model used in the computer simulations, which models how each node in the network decides to make the transition to the new networking technology.  Section~\ref{sec:networks} describes the different network models we looked at and shows the growth curves of how a new networking technology spreads with each network model.  We end the paper in section~\ref{sec:discussion}, where we discuss how the models might apply to a real world system, the limitations of the modeling done here, and some concluding remarks.

\section{Deployment model}
\label{sec:infection}

There are a myriad of factors that affect whether a given node in a network decides to adopt a new networking technology, which makes modeling its spread a very difficult problem.  Our focus here is on how different network models can influence the spread of a new networking technology, however, so we use a very simple deployment model as applied to networks of fixed size.

We also assume that once a node decides to adopt a new networking technology, it will not want to revert to the older technology.  In fact, we assume that there is only one old networking technology and one new networking technology.  To avoid dealing with the influence of converters which provide compatibility between the old and new technologies, we assume a dual-use situation, where nodes with the new technology can continue to use the old technology.

We model the spreading deployment in a network on a fixed graph $G = (V, E)$ representing a network, where $V$ represents the set of nodes and $E$ represents the set of edges between the nodes.  The state of deployment throughout the network changes in one discrete time step, where each node that has not already transitioned to the new networking technology decides whether to transition or not.  We define $D(t)$ to be the set of nodes that have transitioned to the new networking technology so far at time step $t$.

How does a node decide whether to transition or not?  A node would be more inclined to transition if more of its neighbors have already transitioned, since this would allow it to take increasing advantage of the new networking technology to communicate with others.  Contrast this to the case when none of its neighbors have transitioned, in which case the old networking technology would have to be used to communicate with any other node in the network.  A concrete example in the real world would be a single host supporting IPv6 inside an IPv4 subnet, in which case the node cannot use IPv6 to communicate with nodes outside the subnet, ignoring tunneling to simplify matters.

A node would also more inclined to transition if lots of nodes in the network as a whole have already transitioned, sort of not wanting to ``fall behind the times''.  It is unclear how this combines with the previous effect to influence the decision of a node on whether to transition or not, but we will model it as multiplicative rather than additive.  This would have the effect of a node being much more inclined to transition if its neighborhood has already transitioned, given the same general transition rate for the entire network.

On the other hand, transitioning to a new technology is never free, and the cost associated with a transition would discourage a node from making the transition.  We model the cost differently for each network model in section~\ref{sec:networks} to reflect what the nodes and edges represent in the corresponding model, although we do make the simplifying assumption that the costs do not change over time.

Given these factors, we define a utility value $u(v,t)$ which represents how inclined a vertex $v$ is towards making a transition at time step $t$:
\begin{displaymath}
  u(v,t) = \frac{1}{\alpha} \left( h_G(t) h(v, t) - c(v) \right)
\end{displaymath}
where $h_G(t)$ is the total number of transitioned nodes in the network, $h(v,t)$ is the number of nodes adjacent to $v$ that have already transitioned, $c(v)$ is a cost function that is specific to each network model, and $\alpha$ is a scaling factor that is used later:
\begin{displaymath}
 h_G(t) = |D(t)|
\end{displaymath}
\begin{displaymath}
  h(v, t) =| \{ v' \;|\; (v,v') \in E \ \wedge \ v' \in D(t) \} |
\end{displaymath}

\begin{figure}
  \centering
  \includegraphics[width=8cm]{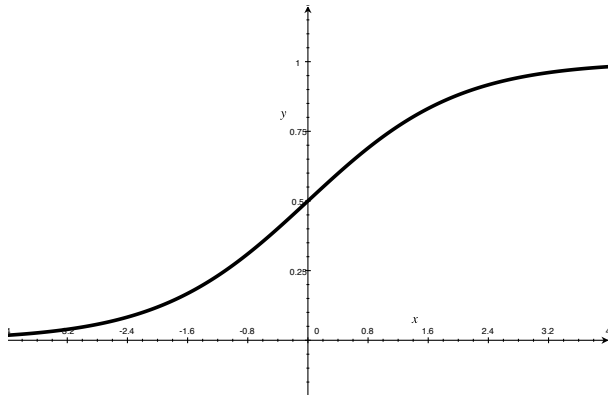}
  \caption{The logistic function, a type of sigmoid function.}
  \label{fig:sigmoid}
\end{figure}

However, a node does not decide whether to transition or not based on whether the utility value is positive or not.  We take advantage of the logistic function, which is graphed out in \figurename~\ref{fig:sigmoid}, to represent the probability that a vertex $v$ will transition or not at time step $t$:
\begin{displaymath}
  p(v,t) = \frac{1}{1+e^{-u(v,t) + \beta}}
\end{displaymath}

Taking advantage of the logistic function ensures that $p(v,t)$ lies between 0 and 1 while having increasing value with increasing utility value.  $\beta$ is a baseline that shifts the logistic function to the right, so that a zero utility value will have low probability.  The scaling factor $\alpha$ in the definition of the utility value is used to ensure that most of the probabilities lie on the center slope rather than near 0 or 1.  Obviously, $u(v,t)$ and $p(v,t)$ do not apply to nodes that have already transitioned, since we assume that nodes do not revert once they make the transition.

The deployment model described in this section is a very simple model of how nodes in a network decide to adopt a new networking technology, but hopefully it captures enough that we can glean insights to how new networking technologies spread given different network models.

\section{Network models}
\label{sec:networks}

In this section, we investigate how different network models influence the deployment of a new networking technology.  The deployment model of section~\ref{sec:infection} is applied to different network models using computer simulations, where we plot out the number of nodes which have made the transition as time goes by.

In the rest of the section, all networks have exactly 10000 nodes.  All of the cases start out with exactly one node having already made the transition, and in most cases we run the simulation until 99\% of the nodes have made the transition.  We fixed $\beta=3$, but the scaling factor $\alpha$ was adjusted for each model to prevent the transition probabilities $p(v,t)$ from clustering around $\frac{1}{1+e^\beta} \approx 0.047$, which would collapse a growth curve to the uninteresting case of section~\ref{sec:independent}.

Because the constants used in the utility functions and transition probabilities are somewhat arbitrary, the number of steps that a network model uses to attain a certain deployment rate is rather arbitrary as well.  This is why we will ignore the absolute speed at which different network models saturate, and instead focus on the features of each deployment growth curve.

\subsection{Independent}
\label{sec:independent}

Before investigating the other network models, we take a look at how a new networking technology would spread if every node decided to make the transition independent of whatever the state of the rest of the network might be.  In other words, we make the exception for the transition probability in the deployment model so that:
\begin{displaymath}
  p(v,t) = \gamma
\end{displaymath}

\begin{figure}
  \centering
  \includegraphics[width=10cm]{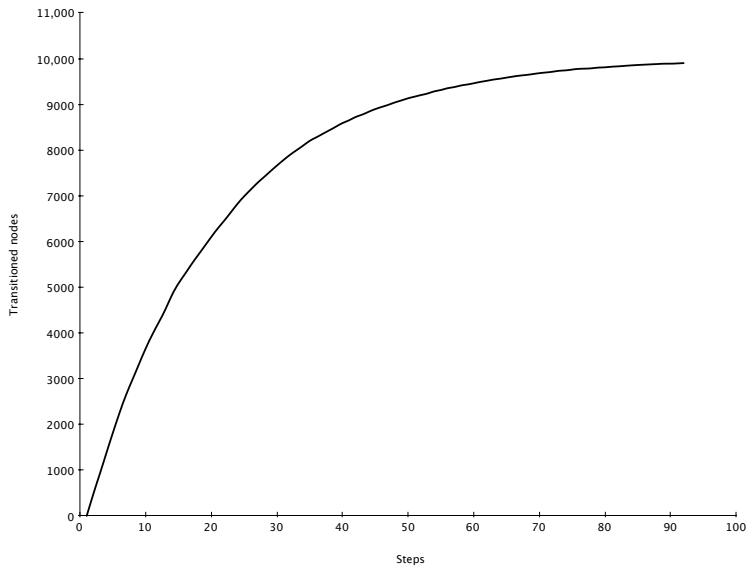}
  \caption{Deployment growth with independent deployment.}
  \label{fig:independent-chart}
\end{figure}

With $\gamma = 0.05$, the growth is shown in \figurename~\ref{fig:independent-chart}, which is unsurprisingly an exponential curve upside down.  This case is included as a comparison for what happens when ``network effects'', the influence of the rest of the network on a single node, are ignored.

\subsection{Clique}
\label{sec:clique}

As an example of a simple network model that still includes ``network effects'', we model the network as a clique, where every node is connected to every other node.  Because the number of transitioned nodes in the entire network is identical to the number of transitioned nodes adjacent to a node that has yet to make the transition, this is basically the case where all of the perceived benefit  is due to how many nodes have made the transition globally.  Because every node is basically the same as every other node, we use a constant cost model:

\begin{displaymath}
  c(v) = \gamma
\end{displaymath}

\begin{figure}
  \centering
  \includegraphics[width=10cm]{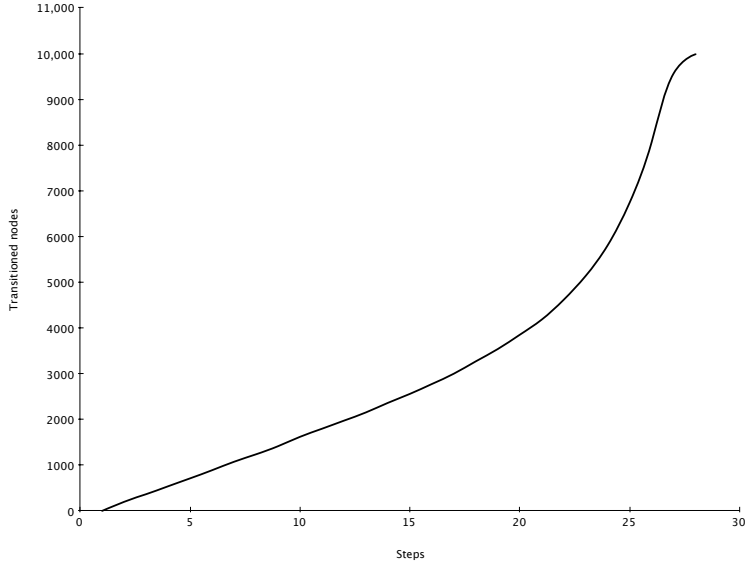}
  \caption{Deployment growth with clique model.}
  \label{fig:clique-chart}
\end{figure}

With $\alpha = \gamma = 1.25 \times 10^7$, the growth for the clique model is shown in \figurename~\ref{fig:clique-chart}.  Deployment steadily grows until it rises exponentially and quickly saturates the network.  This roughly replicates the ``S curve'' that would be expected from technology deployment growth.

\subsection{Random graph}
\label{sec:random}

As a slightly more realistic network model than the clique model of section~\ref{sec:clique}, we apply the deployment model of section~\ref{sec:infection} to the Erd\H{o}s-R\'{e}nyi model~\cite{gilbert:aoms1959} of random graphs.  In this model, every potential edge between nodes in a graph are actualized with probability $p$, which should not be confused with the transition probability $p(v,t)$.  Basically, nodes in the network are randomly connected to each other.

As for the cost model, we assume that it is more expensive to make the transition for nodes that are linked to many nodes.  In addition, we assume that there is a fixed cost for making the transition itself.  Thus we model the cost as follows, where $d(v)$ is the number of edges adjacent to $v$:

\begin{displaymath}
  c(v) = \gamma (1 + d(v))
\end{displaymath}

\begin{figure}
  \centering
  \includegraphics[width=10cm]{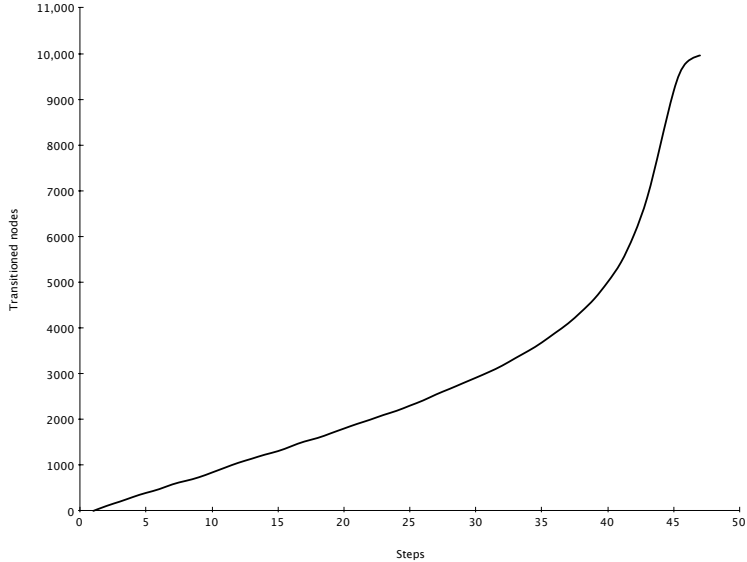}
  \caption{Deployment growth with random graph model.}
  \label{fig:random-chart}
\end{figure}

With $p = 0.001$, $\alpha = 10000$, and $\gamma = 1562.5$, the growth for the random graph model is shown in \figurename~\ref{fig:random-chart}.  An alternate way to interpret the Erd\H{o}s-R\'{e}nyi model is that edges are randomly removed from a clique, so it should be no surprise that \figurename~\ref{fig:random-chart} looks like \figurename~\ref{fig:clique-chart} for the clique model, despite the difference in the cost model.  As in the clique model, deployment steadily grows until it jumps exponentially and quickly saturates the network.

\subsection{Preferential attachment}
\label{sec:preferential}

One possible scenario for how new networking technologies are adopted is that an organization upgrades everything at once, rather than replacing old equipment and software incrementally.  In such a scenario, an organization may be more inclined to make the transition to new networking technology if other organizations it directly communicates with have also made the transition.  The organization would also be more inclined to make the transition if the rest of the world has taken up the new technology.  For example, an ISP may be more inclined to take up IPv6 if most of its peer ISPs have already transitioned to IPv6, and even more so if the rest of the Internet has also taken up IPv6.

In such a scenario, the basic entity that makes a transition would be an organization, and each organization would be influenced by other organizations it has relationships with.  In other words, the network model would be a social network model where vertexes are organizations and edges are relationships.  We will use the Barab\'{a}si-Albert model~\cite{barabasi:science1999} of preferential attachment to model this sort of social network.  In this model, nodes with larger number of neighbors are more likely to collect even more neighbors.

A major organization that has more relationships with other organizations is also likely to have more equipment and personnel, which would raise the cost of making a transition.  Thus we model the transition cost as follows, which also includes a fixed cost:

\begin{displaymath}
  c(v) = \gamma (1 + d(v))
\end{displaymath}

\begin{figure}
  \centering
  \includegraphics[width=10cm]{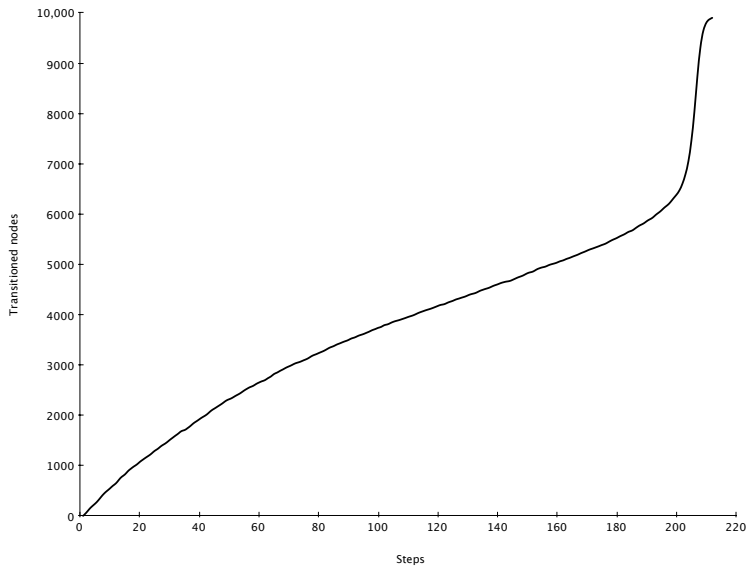}
  \caption{Deployment growth with preferential attachment model.}
  \label{fig:preferential-chart}
\end{figure}

With an initial network of 100 nodes connected as a ring, $\alpha = 3333$, and $\gamma = 2500$, the deployment growth for the preferential attachment model is shown in \figurename~\ref{fig:preferential-chart}.  Superficially, the curve looks like those for the clique and random graph models as shown in \figurename~\ref{fig:clique-chart} and \figurename~\ref{fig:random-chart}, respectively.

A closer look reveals a significant difference, however.  Whereas deployment increases initially with a steady growth rate in \figurename~\ref{fig:clique-chart} and \figurename~\ref{fig:random-chart}, growth slightly \emph{flattens} in \figurename~\ref{fig:preferential-chart} before it skyrockets.  The deployment rate in a preferential attachment model does not quite reflect the traditional ``S curve'' ascribed to technology adoption.  It is not certain why there is a flattening of the growth rate until a critical mass is reached, but it does suggest that caution should be exercised if one attempts to estimate how fast a new networking technology is being widely deployed by fitting a partial growth curve to an ``S curve''.

\subsection{Binary tree}
\label{sec:hierarchical}

Instead of an organization deciding to make the transition to a new networking technology all at once, another scenario would be each piece of equipment being replaced or upgraded one by one.  The decision to upgrade a piece of networking equipment would be influenced by whether neighboring networking equipment has already made the transition, not to mention an industry-wide trend of adopting the new technology.  In this scenario, actual networking hardware would be the vertexes and the communication links between them would be the edges.  A concrete example would be the router-level graph of the Internet.

The router-level graph of the Internet is built up to maximize their utility within the constraints enforced by the hardware technology that is available, which results in a network model that is very unlike a scale-free network constructed through preferential attachment~\cite{alderson:ton2005}.  One of the features which makes them very different is that the routers that form the core of the Internet are typically very high bandwidth, low-degree nodes in the network, since supporting very high bandwidth links makes it extremely difficult to support more than a few links, while the routers with high degree are typically located at the fringe of the network, connecting low-bandwidth end hosts to the rest of the Internet.

While we do not try to realistically model the router-level graph of the Internet here, we will try to model one aspect of the graph: routers near the core are typically a lot more expensive than those at the fringe since they must handle much higher bandwidths.  The network model itself that we use is a very simple binary tree, while the cost to transition a node grows exponentially the nearer it is to the root of the tree.  The cost $c(v)$ is defined as follows, where $l(v)$ is the depth of vertex $v$ in the binary tree:

\begin{displaymath}
  c(v) = \gamma \, 2^{-l(v)}
\end{displaymath}

\begin{figure}
  \centering
  \includegraphics[width=10cm]{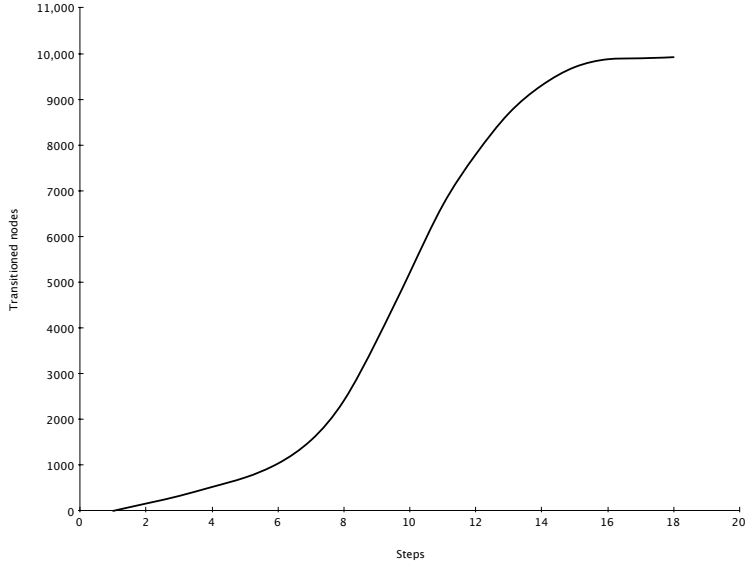}
  \caption{Deployment growth with a binary tree model.}
  \label{fig:hierarchical-chart}
\end{figure}

With $\alpha = 312.5$ and $\gamma =2000000$, the growth for the binary tree model is shown in \figurename~\ref{fig:hierarchical-chart}.  Deployment spreads exponentially at first and then slows down, saturating the network slowly.  This looks more like the traditional ``S curve'' compared to the other models.  The slow saturation at the end seems to be an innate feature of the binary tree model, not an artifact of the constants that we used in the simulation.  In fact, a larger scaling factor $\alpha$ shifts the curve generally to the left, which is basically slowing down the saturation even more.

\begin{figure}
  \centering
  \includegraphics[width=10cm]{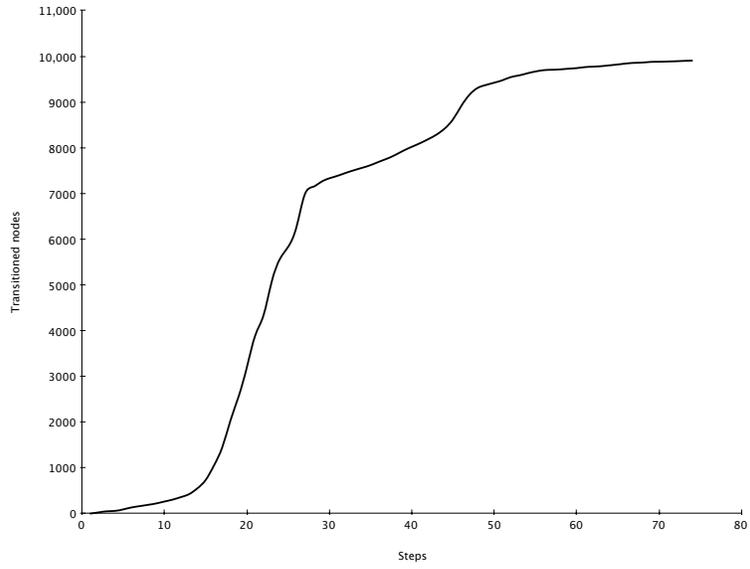}
  \caption{Deployment growth with a binary tree model using a smaller scale factor.}
  \label{fig:hierarchical-chart-scale128}
\end{figure}

Even more interesting behavior occurs when we use smaller scaling factors.  When we set the scaling factor to $\alpha = 78$, we get growth as shown in \figurename~\ref{fig:hierarchical-chart-scale128}.  Instead of a single S-like curve, we get a double-S curve, almost as if there are two growth spurts occuring sequentially.  The smaller scaling factor increases the sensitivity of the deployment model to benefits and costs, so it may be the case that the high transition cost of nodes near the root may delay the deployment in a large subtree in the network, and once the root of this subtree makes its own transition, the rest of the subtree quickly follows.  This could explain the double bursts of growth seen in \figurename~\ref{fig:hierarchical-chart-scale128}.

\begin{figure}
  \centering
  \includegraphics[width=10cm]{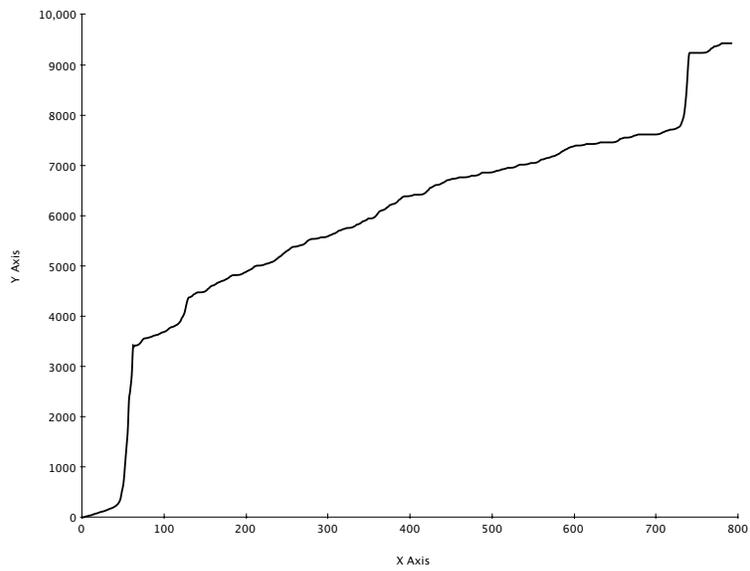}
  \caption{Deployment growth with a binary tree model using a very small scale factor.}
  \label{fig:hierarchical-chart-scale256}
\end{figure}

\begin{figure}
  \centering
  \includegraphics[width=10cm]{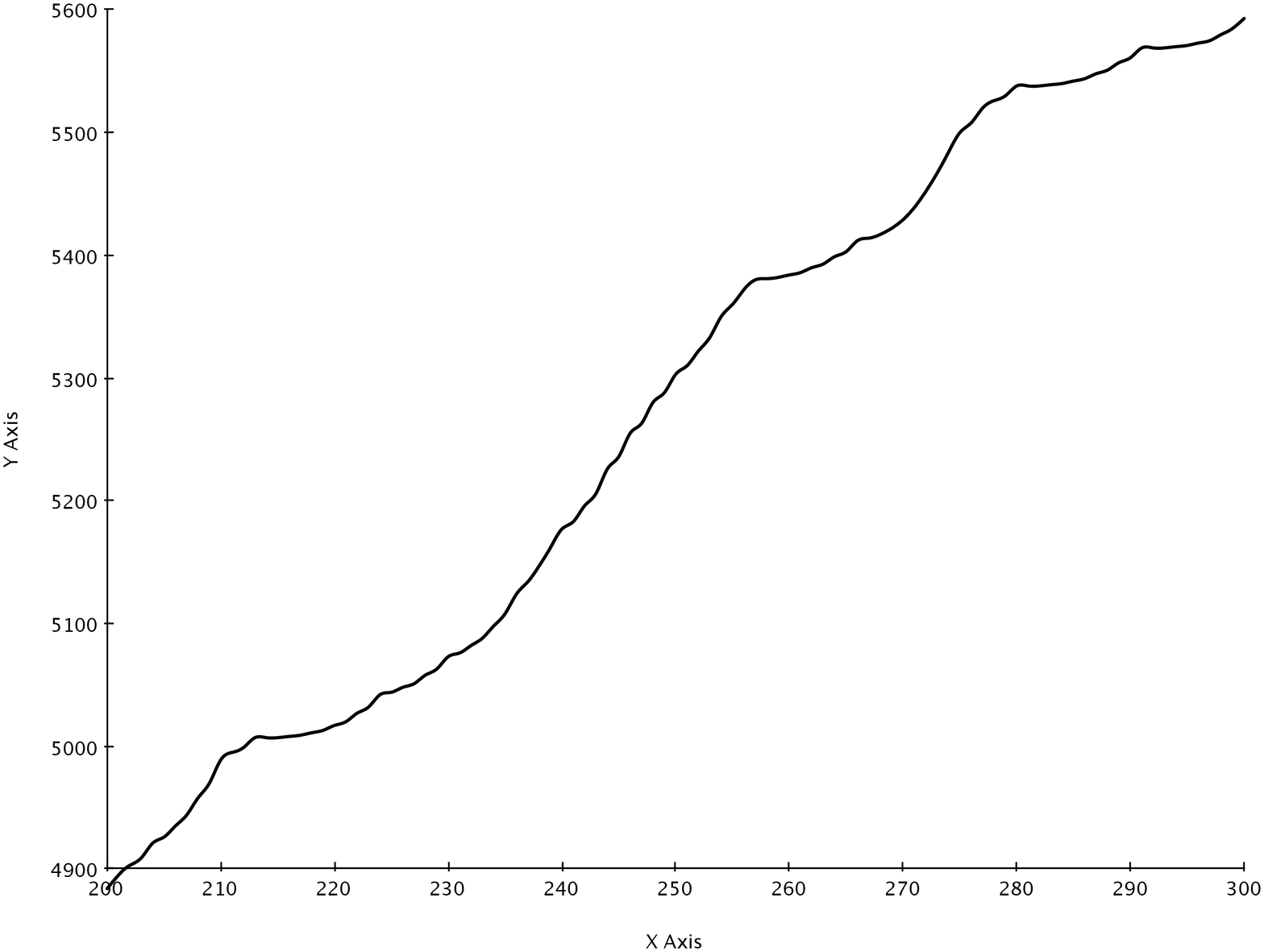}
  \caption{A zoomed view of \figurename~\ref{fig:hierarchical-chart-scale256}.}
  \label{fig:hierarchical-chart-scale256-zoom}
\end{figure}

Yet smaller scaling factors, which is equivalent to an increased cost-benefit sensitivity, gives rise to even more complex growth curves.  When we set the scaling factor to $\alpha = 39$, we get the growth curve shown in \figurename~\ref{fig:hierarchical-chart-scale256}, where we can see many growth bursts occuring throughout time.  The curve shows characteristics of a self-similar curve, which is made clearer when we zoom into a small portion of the curve as in \figurename~\ref{fig:hierarchical-chart-scale256-zoom}.

While the binary tree model is still an unrealistic model for real networks, the unexpected behavior in the growth curve shows how different network and cost models can result in qualitatively different behavior in the spread of new networking technologies.

\section{Discussion}
\label{sec:discussion}

\subsection{Applicability to real networks}
\label{sec:applicability}

One obvious real-world network we may want to compare the network models in section~\ref{sec:networks} with is the spread of IPv6 in a world of IPv4.  Unfortunately, this is difficult not only because IPv6 deployment is still in its initial stages, but also because actually measuring how widely IPv6 is deployed is a very hard problem.  Still, we will briefly discuss how the models compare with the growth of IPv6 address block allocations to the Regional Internet Registries, which can be found in \figurename~2 of \cite{karpilovsky:pam2009}.

If one were to look at \figurename~2 of \cite{karpilovsky:pam2009}, one might notice that the growth curves for RIPENCC and APNIC are vaguely similar to the growth curve for the preferential attachment model in \figurename~\ref{fig:preferential-chart}.  They start off with the growth rate slowly flattening until they begin to rise again.  The growth for the other regions look more like the other models, although there is not enough data to see if the growth curves look more like one model rather than another.

For RIPENCC and APNIC, the similarity of the growth curves to \figurename~\ref{fig:preferential-chart} might suggest that organizations are transitioning all at once to IPv6 rather than incrementally, at least in terms of allocating IPv6 address blocks, and that they are influenced by other organizations having allocated IPv6 address blocks of their own.  Or perhaps the similarity is just a coincidence, the initial growth spurt being due to a clarification of allocation policy which gradually tapered off as suggested in \cite{karpilovsky:pam2009}, and the later rise in growth rates being due to the impending depletion of IPv4 addresses.

Regardless, the similarities are a tantalizing hint that network structure can indeed influence the spread of a new networking technology in a real world situation, and in fact, may be an indication of the decision processes by which organizations determine their plans concerning the deployment of IPv6.  This may be an issue that deserves a closer look.

\subsection{Limitations}
\label{sec:limitations}

Both the deployment model in section~\ref{sec:infection} and network models in section~\ref{sec:networks} are very limited in their realism.  In the following, we discuss what those limitations are, which also point to directions for further research.

\begin{enumerate}
\item Networks grow.  We have assumed networks of fixed size, which does not reflect the reality that even incumbent networks such as IPv4 are growing.  Similarly, the way we model deployment would be like embedding an IPv6 network inside a fixed IPv4 network, and technically, an IPv6 network is entirely separate from an IPv4 network.  Dual-stack IPv6/IPv4 deployments may be the rule while IPv4 dominates, but this is unlikely to continue to be the case once IPv6 becomes widespread.
\item The decision process for whether a node will make the transition or not should be modeled more realistically.  There will be many more factors that influence the decision process.  For example, the increasing scarcity of IPv4 addresses along with growth would presumably encourage the deployment of IPv6.  In addition, constants used in an ideal deployment should be based on measurable values that are dependent on the underlying factors, rather than being derived by fitting to a curve.
\item The network models should be more realistic.  For example, we may look at actual peering relationships between ISPs to model the relationships between organization.  We could also use more realistic models of communication networks such as heuristically optimal topologies~\cite{alderson:ton2005}.
\item A look at how new networking technologies spread throughout a network.  We have only looked at the total number of deployments in a network, and seeing which nodes actually make the transition as time passes would undoubtedly give rise to new insights.  Visualizing this in a large network is likely to be a challenging problem.
\item Actual data to validate models would be very useful.  As can be seen in the measurement of IPv6 deployment rates, defining what exactly a deployment means can be ambiguous, and actually measuring the deployment rate can be very difficult even when it is defined.
\end{enumerate}

\subsection{Concluding remarks}
\label{sec:conclude}

With a simple model for how new networking technologies get deployed throughout a network and applying it to different network models, we have seen how different network structures and cost models can change the qualitative behavior of how a new networking technology spreads.  While the research is in its infancy such that we are not able to make predictions or derive insights from existing data with any certainty, we can make the observation that any realistic model of how new network technologies get deployed will have to account for network structure and related cost models.

\bibliography{strings,articles,web,proceedings}
\bibliographystyle{plainnat}

\end{document}